\documentclass{article}

\font\scap=cmcsc10

\def\smallskip{\vskip 3pt}
\def\medskip{\vskip 6pt}
\def\bigskip{\vskip 12pt}

\newcount\eqnumber
\def\neweq{{\rm{(\the\eqnumber)}}\global\advance\eqnumber by 1}
\def\eqdef#1{\eqno\xdef#1{\the\eqnumber}\neweq}
\def\newaeq{{\rm{(\the\eqnumber a)}}\global\advance\eqnumber by 1}
\def\eqdaf#1{\eqno\xdef#1{\the\eqnumber}\newaeq}
\def\eqdisp#1{\xdef#1{\the\eqnumber}\neweq}
\def\eqdasp#1{\xdef#1{\the\eqnumber}\newaeq}

\newcount\refnumber
\def\newref{{\the\refnumber}\global\advance\refnumber by 1}
\def\refdef#1{{\xdef#1{\the\refnumber}}\newref}

\font\tenmsb=msbm10
\font\sevenmsb=msbm7
\font\fivemsb=msbm5
\newfam\msbfam
\textfont\msbfam=\tenmsb
\scriptfont\msbfam=\sevenmsb
\scriptscriptfont\msbfam=\fivemsb

\def\misi{\raise.5ex\hbox{$\scriptscriptstyle 1$}\hbox{\kern-.12em$\scriptscriptstyle /$}\kern-.12em\lower.5ex\hbox{$\scriptstyle 2$}}
\def\miso{\raise.5ex\hbox{$\scriptstyle 1$}\hbox{\kern-.12em$\scriptstyle /$}\kern-.12em\lower.5ex\hbox{$\scriptstyle 2$}}

\usepackage{graphicx}
\usepackage{bxcjkjatype}

\def\ten{taishi }
\def\tan{taishi}

\newcommand{\lsqbr}{\big[}
\newcommand{\rsqbr}{\big]}

\begin{document}

\centerline{\bf On the singularities of the discrete Korteweg-deVries equation}
\bigskip

{\scap  Doyong Um}\quad
{\sl Department of Mathematics, The University of Auckland, 38 Princes Street, Auckland, 1010, New Zealand}
\medskip{\scap A. Ramani} and {\scap B. Grammaticos}\quad
{\sl  Universit\'e Paris-Saclay, CNRS/IN2P3, IJCLab, 91405 Orsay, France } 
and {\sl  Universit\'e de Paris, IJCLab, 91405 Orsay France}
\medskip {\scap R. Willox} \quad
{\sl Graduate School of Mathematical Sciences, the University of Tokyo, 3-8-1 Komaba, Meguro-ku, 153-8914 Tokyo, Japan }
\medskip{\scap J. Satsuma}
\quad{\sl  Department of Mathematical Engineering, Musashino University, 3-3-3 Ariake, Koto-ku, 135-8181 Tokyo, Japan}

\bigskip
{\sl Abstract}
\smallskip
We study the structure of singularities in the discrete Korteweg-deVries (d-KdV) equation. Four different types of singularities are identified. The first type corresponds to localised, `confined', singularities, the confinement constraints for which provide the integrability conditions for generalisations of d-KdV. Two other types of singularities are of infinite extent and consist of oblique lines of infinities, possibly alternating with lines of zeros. The fourth type of singularity corresponds to horizontal strips where the product of the values on vertically adjacent points is equal to 1. (A vertical version of this singularity with product equal to $-1$ on horizontally adjacent sites also exists). Due to its orientation this singularity can, in fact, interact with the other types. This leads to an extremely rich structure for the singularities of d-KdV, which is studied in detail in this paper. Given the important role played by the fourth type of singularity we decided to give it a special name: {\sl \tan} (the origin of which is explained in the text).
The \ten do not exist for nonintegrable extensions of d-KdV, which explains the relative paucity of singularity structures in the nonintegrable case: the second and third type of singularities that correspond to oblique lines still exist and the localised singularities of the integrable case now become unconfined, leading to semi-infinite lines of infinities alternating with zeros.

\bigskip
1. {\scap Introduction}
\medskip
The study of the singularities of a system is often the first step towards the assessment of its integrable, or nonintegrable, character. In the case of differential equations the existence of singular points (points in the neighbourhood of which the solution is not analytic) may interfere with the construction of a global solution by analytic continuation. This is true in particular when in
the neighbourhood of a singular point {the solution becomes multivalued}. In order to be able to define the solution of a differential equation one must be able to do away with multivaluedness, the difficulty being however that {nonlinear systems may possess movable singularities, i.e. singularities the location of which} depends on the initial conditions. In this case the uniformisation methods which work for linear differential equations can no longer be applied. The solution to this gordian knot was given by Painlev\'e who decided that a global solution of a given nonlinear differential equation could only exist if movable multivaluedness-inducing singularities were absent  [\refdef\painleve]. This {requirement} came to be known as the Painlev\'e property and the widely accepted conjecture is that the solutions of nonlinear differential equations integrable through spectral methods {all} possess this property. While Painlev\'e's approach was tailored to ordinary differential equations, it proved possible to extend the notions of singularity and singlevaluedness to the case of partial differential equations, once a critical mass of results on integrable multidimensional systems was reached [\refdef\ars].

The situation for discrete systems is more complicated. First, one must decide what {constitutes a singularity and what the equivalent of the Painlev\'e property is for a discrete system.}  And then one must question the practice, which is common in the case of differential systems, to consider the Painlev\'e property as synonymous to integrability. In order to make {these difficulties and the solutions they require intelligible} to our readers, let us consider the simple case of second-order mappings. 
Suppose that, due to a special choice of initial conditions, at some iteration step a degree of freedom is lost. In the case at hand this would mean that the value of the dependent variable at point $n+1$ does not depend on the value at point $n-1$ (and in fact that the inverse mapping is not defined at point $n+1$). This loss of degree of freedom constitutes a singularity for the mapping. Now, in most cases when a degree of freedom is lost, it is never recovered. However there exist cases where iterating past a singularity one arrives after a finite and usually small number of steps at an indeterminate result. The way to lift this indeterminacy is through continuity with respect to the initial conditions, typically by introducing a small quantity and taking the limit where this quantity goes to zero. {If this procedure allows to lift the singularity} and the degree of freedom is actually recovered, we say that the singularity is confined [\refdef\sincon]. Our contention is that the singularity confinement property constitutes the discrete analogue of the Painlev\'e property. The evidence in support of this claim is that all known discrete systems which are  integrable through spectral methods do possess the singularity confinement property. On the other hand, most systems known to be nonintegrable have unconfined singularities. However there exist systems which, despite being nonintegrable,  do possess confined singularities. This calls for caution in the use of singularity confinement as an integrability criterion, but we shall not go into further details here [\refdef\desoto,\refdef\redemp]. 

{Let us rather illustrate our arguments with some concrete examples which will be helpful for understanding the body of this paper. Take for example the} nonintegrable mapping (where $a\neq0$)
$$x_{n+1}+x_{n-1}=a+{1\over x_n^2}.\eqdef\zena$$
A singularity appears when due to the choice of initial conditions the value of {$x$ becomes 0 at some step $n$.} Iterating further we find the succession of values
$$\{0,\infty^2, a,\infty^2, 0, \infty^2, a,\infty^2, 0,\cdots\},$$
and the singularity does not disappear: it is unconfined. (The meaning of the notation $\infty^2$ is the following: had we introduced a small quantity $\epsilon$ instead of 0 for the value of $x_n$, we would have found that the dominant behaviour of the term in question would have been $1/\epsilon^2$). 

Now if we consider the integrable mapping (with $a\neq0$)
$$x_{n+1}+x_{n-1}=a+{1\over x_n},\eqdef\zdyo$$
we find that the singularity that starts from 0 disappears after five iteration steps. The corresponding pattern is 
$$\{0,\infty, a,\infty, 0\}.$$
This singularity is confined. However another type of singularity does exist. If at some iteration of (\zdyo), or in fact of (\zena), the value of $x$ is infinite we obtain the succession 
$$\{\cdots,\infty, f,\infty, f', \infty, f'',\cdots\},$$
where $f,f',f'',\cdots$ are finite values depending on the initial condition. { As the pattern $\{\infty, f\}$ repeats indefinitely we call this type of singularity `cyclic' [\refdef\express]. Cyclic singularities may  exist for integrable as well as for non-integrable mappings and their presence has no bearing on the issue of integrability or nonintegrability of a given mapping  [\refdef\halburd].}

{This however does not exhaust all possible types of singularities encountered in second-order mappings.} Consider the (integrable) mapping
$$(x_{n+1}+x_n)(x_n+x_{n-1})=a(x_n^2-1).\eqdef\ztri$$
It has two confined singularity patterns,  $\{1,-1\}$ and  $\{-1, 1\}$, but also the pattern
$$\{\cdots,\infty,\infty,\infty,x_0,-x_0,\infty,\infty,\infty,\cdots\},$$
{in which the singularities extend indefinitely both ways from a finite set of regular values. This is an example of a singularity  of a  type that we have dubbed `anticonfined' [\refdef\anticon].} 
Anticonfined patterns can also exist for nonintegrable mappings. For instance the mapping
$$x_{n+1}=x_{n-1}\left(x_n-{1\over x_n}\right),\eqdef\ztes$$
has the anticonfined pattern
$$\{\cdots,0^5,0^3,0^2,0,0,x_0,0,\infty,-x_0,\infty,\infty,\infty^2,\infty^3,\infty^5,\cdots\},$$
where the exponents  in the pattern grow following the Fibonacci sequence. As we have explained in [\anticon], an exponential growth of the exponents in an anticonfined pattern is an indication of nonintegrability, while polynomial growth is compatible with a possible integrable character of the mapping. 

How does all this translate to the case of lattice equations? It turns out that not much is known concerning the singularities of multidimensional systems (hence the interest of the present work). In fact, even for the discrete KdV equation which is the paradigmatic lattice integrable system, the structure of its singularities has never been studied in depth. (Possible singularities in type-Q equations in the ABS-classification [\refdef\abs] are investigated in [\refdef\atkin] from the point of view of multidimensional consistency, but the singular values that may appear in non type-Q equations like the d-KdV equation  present particular difficulties, as explained in [\refdef\kinson]).  

In [\sincon], which is the paper that introduced the notion of singularity confinement, some of the present authors studied the singularity of the discrete KdV equation [\refdef\hirota]
$$x_{m+1,n+1}=x_{m,n}+{1\over x_{m+1,n}}-{1\over x_{m,n+1}},\eqdef\zpen$$
{when a 0 appears at some vertex  as a consequence of the initial conditions, in the case where these are given on a staircase.} It was shown there that the singularity is confined to one elementary cell of the lattice, while it propagates indefinitely if (\zpen) is altered slightly so as to become nonintegrable. The existence of more complicated singularity structures was acknowledged in that paper but these did not make the object of a specific study. The question of the singularities of the discrete KdV equation was addressed again recently in [\refdef\doyong] by the authors of the present paper. This recent work was inspired by the ARS approach [\ars] to evolution equations, which consists in studying all possible one-dimensional reductions of a two-dimensional system. In this spirit we considered reductions of the type $x_{m+1,n}=x_{m,n+q}$ and studied the singularities of the resulting mapping. One important finding of that study was the discovery that singularities of anticonfined type do exist for the reduced system. Still that study did only scratch the surface as far as the classification of singularities for the full d-KdV equation is concerned. 
In this paper we intend to remedy this situation by studying in detail the possible singularities of the discrete KdV and the way in which they interact.
\bigskip
2. {\scap The typology of singularities of discrete KdV}
\medskip
In [\sincon] a first type of singularity of the discrete KdV was identified. The simplest realisation of such a singularity is shown in Figure 1. The singular values occupy the vertices of an elementary cell and the singularity is confined, meaning that it is surrounded by finite values {(not shown in the figure).} 
\bigskip
\centerline{\includegraphics[width=4.5cm,keepaspectratio]{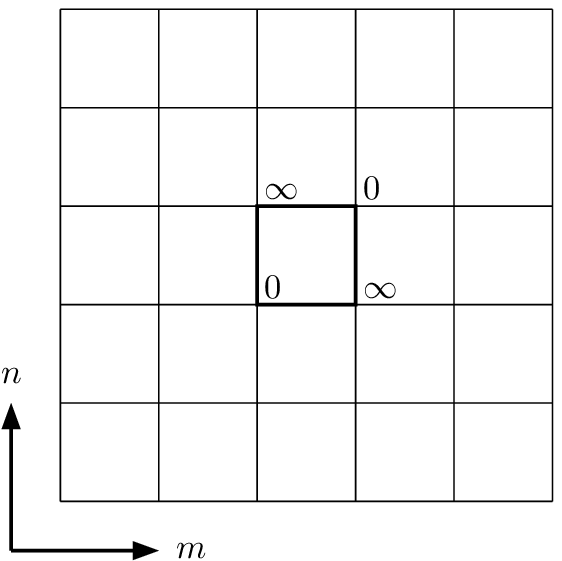}}
\vskip-.4cm\centerline{~~~~{\sl Figure 1}}\vskip.5cm

This is by far not the only singularity of this type. When two zeros appear on neighbouring vertices {on a staircase} as shown in Figure 2 (inner part), the singularity extends over a cross-shaped domain before being confined. More complicated patterns are obtained when one considers more than two zeros situated on neighbouring vertices, as seen {on the south-west edge} of Figure 2, where we have four neighbouring zeros. One obtains a rhombus-like shape for the domain {that contains all singular} values, with horizontal and vertical extension $2N-1$, where $N$ is the number of neighbouring zeros in the initial conditions. \bigskip
\bigskip
\centerline{\includegraphics[width=8cm,keepaspectratio]{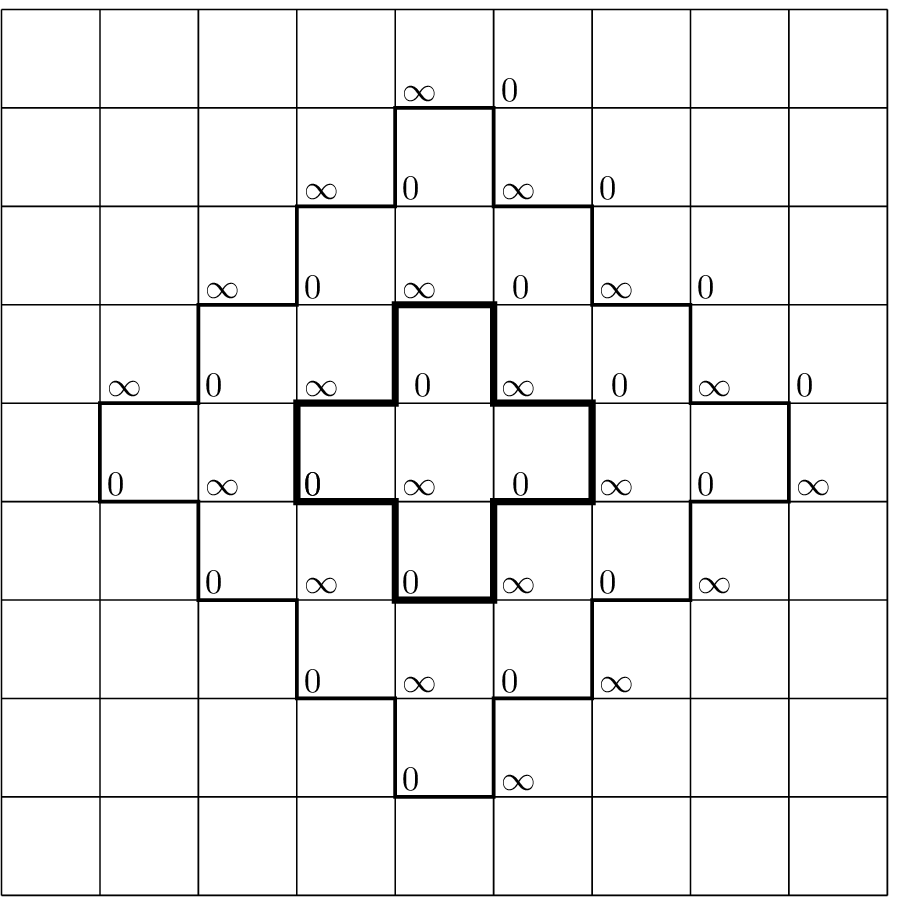}}
\vskip.25cm\centerline{~~~~{\sl Figure 2}}\vskip.5cm

Since there is no restriction on the initial condition one may consider, the value of $N$ can be arbitrarily high, which means that the domain occupied by singular values can be arbitrarily extended. {This does not pose any problems} as far as the confined character of the singularity is concerned. Since the extension of the domain of the singular values is limited by the number of neighbouring zeros, one always has a confined singularity unless the number of zeros was already infinite in the initial condition [\doyong].  

A second type of singularity can be obtained by simple inspection of (\zpen). Taking for $x_{m,n}$ an infinite value results in an infinite value for $x_{m+1,n+1}$. However iterating (\zpen) backwards, i.e. in the south-west direction, leads also to an infinite value for $x_{m-1,n-1}$. Thus we obtain a singularity which extends indefinitely on a diagonal along the SW/NE direction, i.e. $x_{m+k,n+k}=\infty$ for arbitrary integer $k$. Figure 3 depicts such a situation (on the left). {It is important to stress however that this situation does not correspond to an unconfined singularity as} one cannot enter this singularity coming from regular values. Since all such lines of infinities run in the same direction in the $(m,n)$ plane one can of course have an arbitrary number of them separated, or not, by strips of regular values. 

A third type of singularity is also one of infinite extent. It consists of {an arbitrary number of} parallel lines running towards the NE-direction, just as in the previous case, but the values alternate between infinity and zero {in the horizontal (or vertical) direction}, with infinities {necessarily} appearing on the two external lines. The simplest case of just three lines is presented on the right in  Figure 3. Again this singularity does not create a conflict with confinement since it is of infinite extent in both directions and does not contain any regular values so that it cannot be entered into.
\bigskip
\centerline{\includegraphics[width=7cm,keepaspectratio]{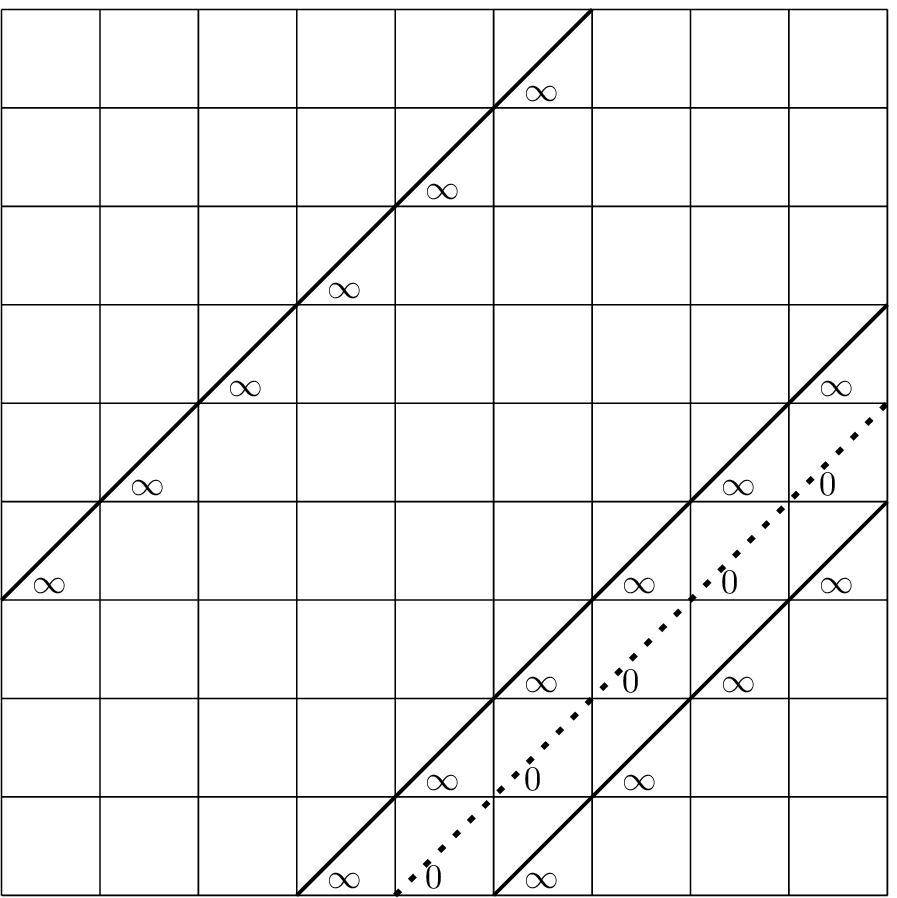}}
\vskip.25cm\centerline{~~~~{\sl Figure 3}}\vskip.5cm

{In order to understand the origin of the fourth and final type of singularity,} we start from (\zpen) and rewrite it as
$${x_{m+1,n+1}x_{m+1,n}-1\over x_{m+1,n}}={x_{m,n}x_{m,n+1}-1\over x_{m,n+1}}.\eqdef\zhex$$
We readily see that if at some point $(m,n)$ the product $x_{m,n}x_{m,n+1}$ is equal to 1 then it will stay equal to 1 at the next point on  the right, i.e. $x_{m+1,n}x_{m+1,n+1}$ will also be equal to 1. Since this is valid in both the forward and backward directions this singularity consists of a horizontal strip, running from west to east, where at each vertical {we have two reciprocal values. }Such a situation is represented in the upper part of Figure 4. 
\bigskip
\centerline{\includegraphics[width=6cm,keepaspectratio]{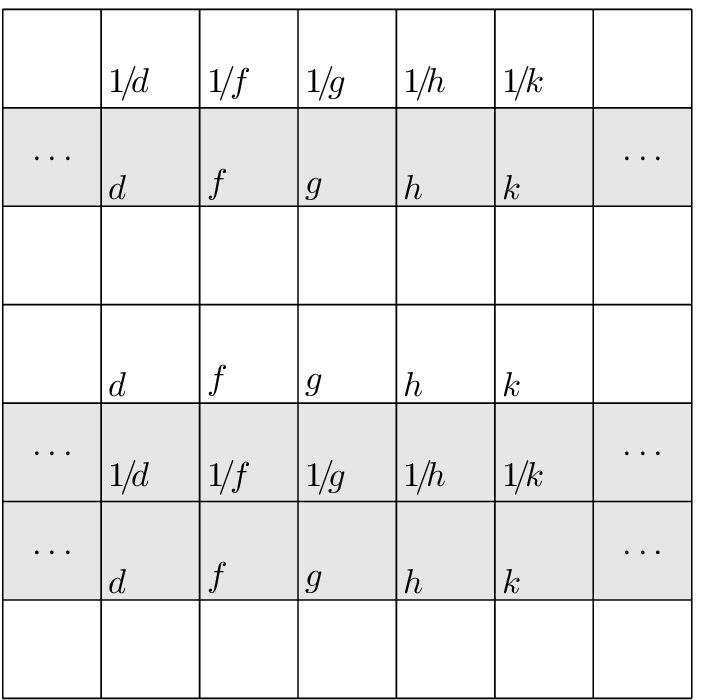}}
\vskip0.1cm\centerline{~~~~{\sl Figure 4}}\vskip.5cm

{Moreover, calculating e.g. $x_{m+1,n+2}$ from (\zpen) when $x_{m,n}x_{m,n+1}=1$,
$$\displaylines{\hskip4cm x_{m+1,n+2} = x_{m,n+1} + {1\over x_{m+1,n+1}} - {1\over x_{m,n+2}} \hfill\cr
\hskip5.55cm = {1\over x_{m,n}} + {x_{m+1,n}} - {1\over x_{m,n+2}} \hfill\cr
\hskip5.55cm  = {1\over x_{m,n}} + \big( x_{m,n-1} + {1\over x_{m+1,n-1}} - {1\over x_{m,n}}\big) - {1\over x_{m,n+2}} \hfill\cr
\hskip5.55cm  = x_{m,n-1} + {1\over x_{m+1,n-1}} - {1\over x_{m,n+2}},\hfill\cr}$$


one sees that its value does not depend on $x_{m,n}$ at all and, in fact, that the same is true for any lattice site not contained in the horizontal strip. This means that, outside the strip, this degree of freedom has been irretrievably lost and that such a horizontal strip is indeed a singularity for d-KdV. However, as this singularity extends over a whole strip, effectively, it cannot be entered into spontaneously and it does not impede the integrability of d-KdV.}

Clearly one can imagine situations where more than one such strip exists, and even cases where several of them are adjacent. The lower part of Figure 4 depicts such a situation showing a `double' strip, i.e. one where we have $x_{m,n}x_{m,n\pm1}=1$. It goes without saying that we could have reorganised equation (\zpen) in a different way {to deduce the existence of vertical strips} with $x_{m,n}x_{m+1,n}=-1${, with properties similar to those of the horizontal ones}. 

{In the following we shall see that although these singularities are completely `transparent' vis-a-vis the evolution of regular values on the lattice, they nonetheless play a vital role in the analysis of the singularities for d-KdV as they can interact with singularities of the other types.} {In the following section, we shall only describe the interaction properties of singularities consisting of horizontal strips, such as in Figure 4, but the case of vertical strips can be easily deduced, mutatis mutandis, from the properties of the horizontal ones.}

\bigskip
3. {\scap  The interaction of singularities}
\medskip
The first three types of singularities described in the previous section {can} co-exist on the lattice without interfering with each other. (A possible exception could have been the case where an infinity on a vertex of a singularity of the first type happens to lie on the line of infinities of a singularity of the second or third type but this turns out to be a case of a different interaction, as we shall explain below). The only remaining possibility is thus an interaction of a singularity of the fourth type with one of the other types. 

{In the following we choose to evolve (\zpen) in the positive $m$ and $n$ directions, as indicated in Figure 1. Let us start with the simplest case, that of an interaction of a horizontal strip such as in Figure 4 with an elementary cell singularity.} The result of this interaction depends on where the strip meets the elementary cell. If the interaction takes place ``head-on'' then, essentially, nothing happens. The upper part of Figure 5 represents such a situation. 
\bigskip
\centerline{\includegraphics[width=5.5cm,keepaspectratio]{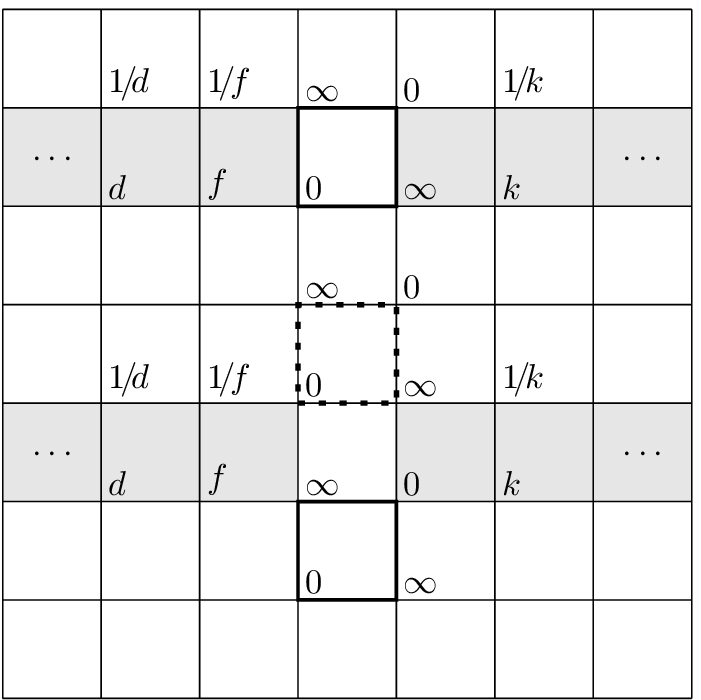}}
\vskip.25cm\centerline{~~~~{\sl Figure 5}}\vskip.5cm

However if the strip  meets the upper part of the elementary cell {(in bold in the lower part of figure 5)} this is different: the interaction creates a kind of {mirage} of the initial elementary cell {(in dotted lines)} but the {strip traverses} the interaction zone {unperturbed}. The same situation can be observed when such a strip interacts with a more complicated localised singularity, like the rhombus-shaped one we met in section 2. Again a head-on interaction leaves both singularities unperturbed but an off-centre interaction leads to the creation of a shifted image of the initial rhombus, while the strip of inverse values itself is unperturbed. Figure 6 depicts such a situation. 
\bigskip
\centerline{\includegraphics[width=7cm,keepaspectratio]{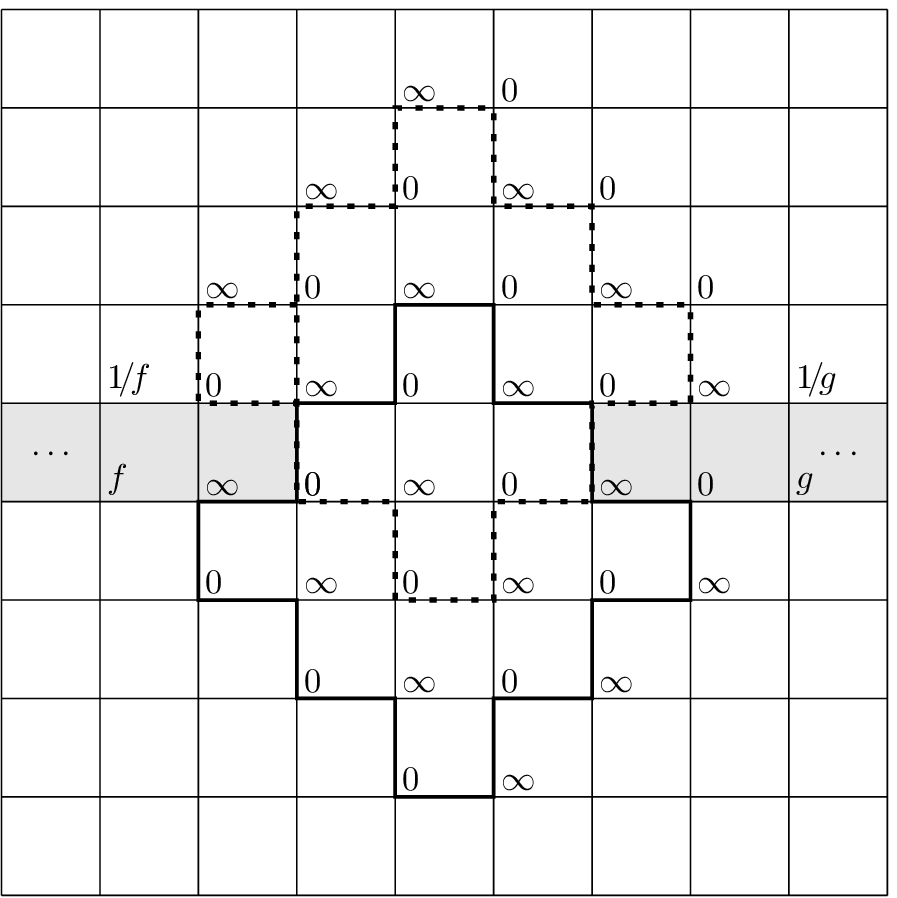}}
\vskip.25cm\centerline{~~~~{\sl Figure 6}}\vskip.5cm

{At this point is natural to ask what happens when such a strip meets the lower part of an elementary cell.} It turns out that this question is {best answered by considering the interaction of a strip with a SW/NE diagonal of infinities.} The result of such an interaction is presented in Figure 7: both the line of infinities as well as the horizontal strip  jump two lattice spacings upwards. 
\bigskip
\centerline{\includegraphics[width=6cm,keepaspectratio]{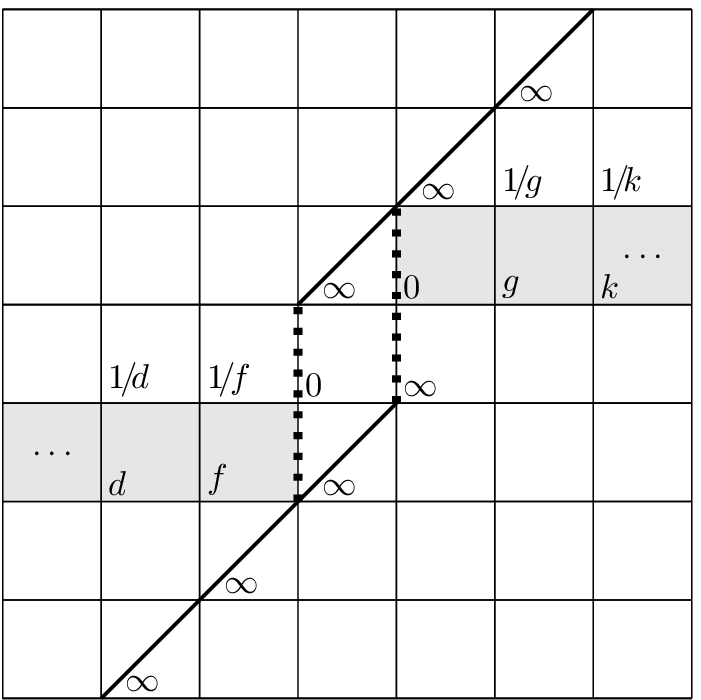}}
\vskip.25cm\centerline{~~~~{\sl Figure 7}}\vskip.5cm

However this is not the only possible interpretation of this interaction. Just as in {astronomy where constellations are defined by arbitrarily drawing lines that link the stars so as to conform to a chosen interpretation, we could have represented the interaction region as in Figure 8, where after interaction the line of infinities is shifted towards the north-west while the strip  jumps two lattice spacings upwards.} {Note that the center of this interaction (the dotted lines in Figure 8) is nothing but the elementary cell of singularities. As it is impossible for a horizontal strip to meet the lower part of such a cell without a diagonal of singularities extending to the south-west, this answers our question.
\bigskip
\centerline{\includegraphics[width=6.cm,keepaspectratio]{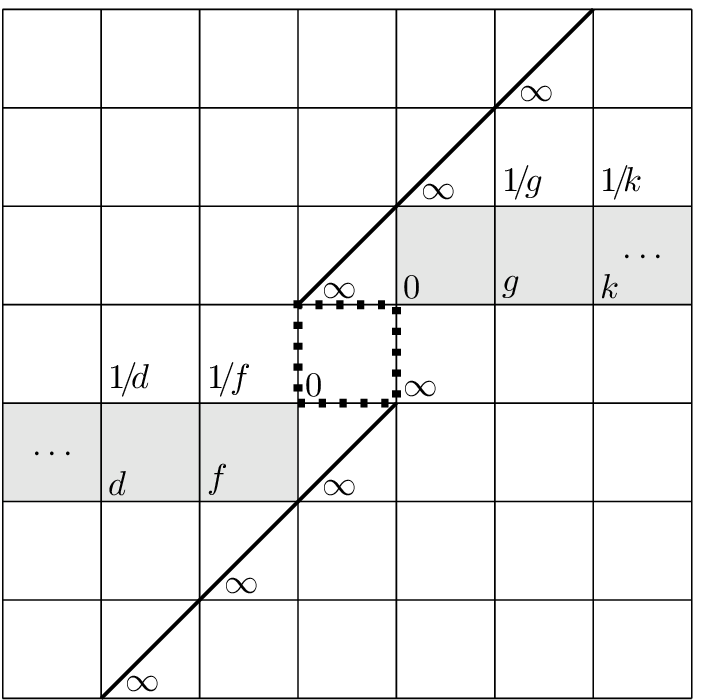}}
\vskip.25cm\centerline{~~~~{\sl Figure 8}}\vskip.5cm
{In the more complicated situations that will be described below we shall always draw the lines in the interaction region according to the convention chosen for Figure 7, rather than that in Figure 8.}

If we start with {two lines of infinities, side by side,  the end result  (shown in Figure 9)} is what one would expect from the independent interaction of the strip with each {consecutive} line of infinities.

\bigskip
\centerline{\includegraphics[width=7cm,keepaspectratio]{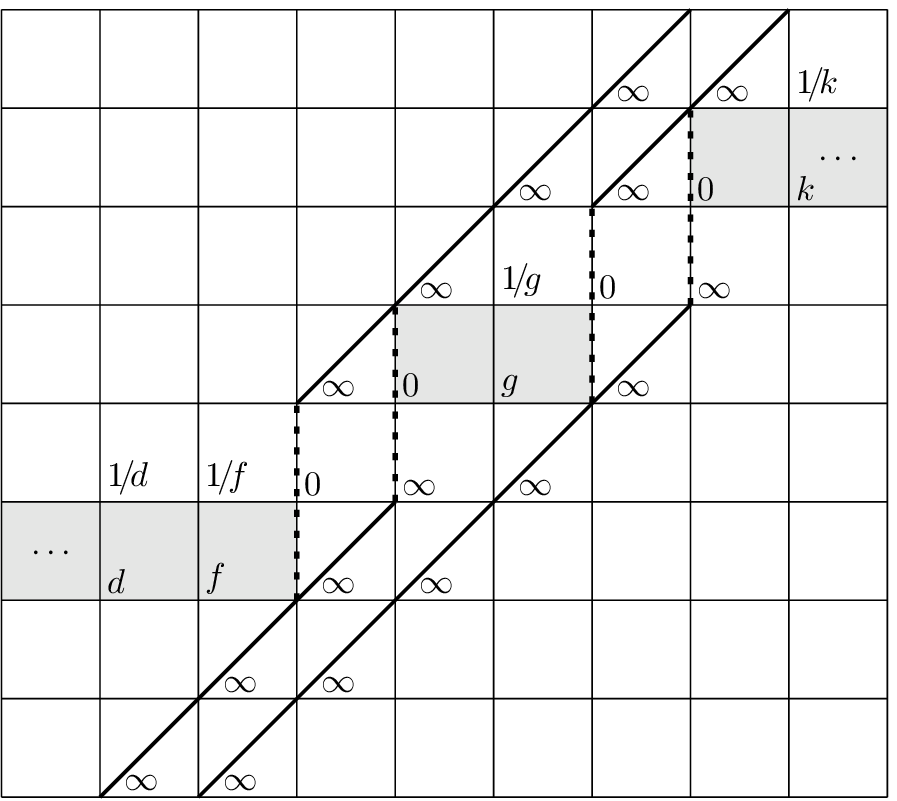}}
\vskip.25cm\centerline{~~~~{\sl Figure 9}}\vskip.5cm

A similar structure can be observed in Figure 10 in which the horizontal strip interacts with a singularity of the third type, i.e. parallel lines with alternating infinities and zeros. 
\bigskip
\centerline{\includegraphics[width=7cm,keepaspectratio]{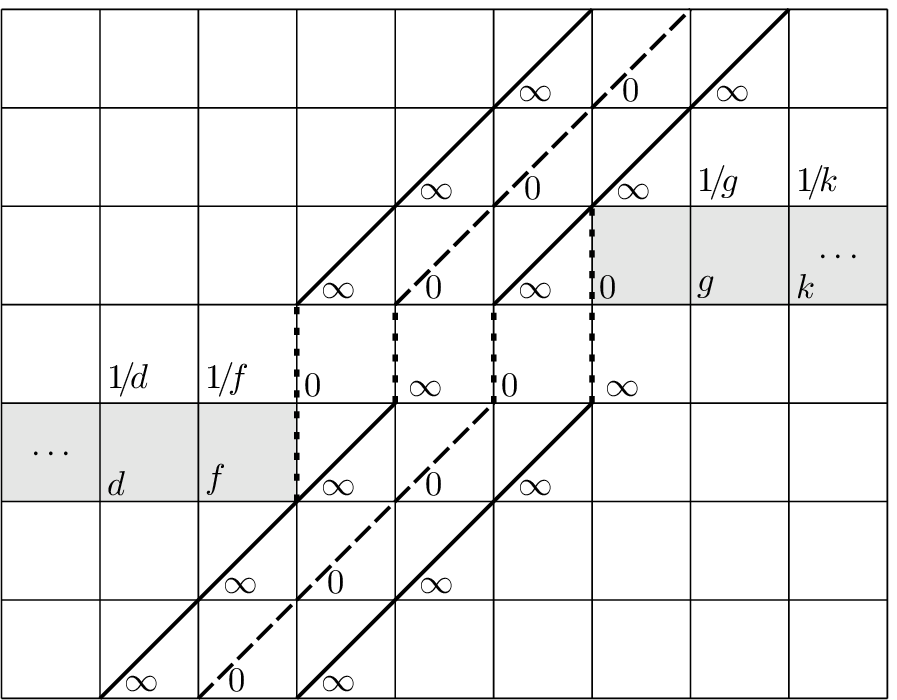}}
\vskip.25cm\centerline{~~~~{\sl Figure 10}}\vskip.5cm

{All results we presented up to now were obtained by taking initial conditions in which a small quantity $\epsilon$ is introduced, and then by taking the limit $\epsilon\to0$.} The various zeros appearing in the figures are thus proportional to $\epsilon$, the infinities to $1/\epsilon$ and the values on the horizontal strip are such that their respective products are equal to $1+{\cal O}(\epsilon)$. So our analysis up to this point has assumed tacitly that all ``zeros" appearing in the various singularities are of the same order and that the product of any of the ``infinities'' with any of the zeros would lead to a finite number. {This is too restrictive an assumption and it should be relinquished if one wishes to discover the unfettered richness of the singularities of discrete KdV.  We shall therefore assume that the line of infinities is obtained as the limit of  $1/\epsilon^q$, for some positive integer $q$,  while the inverse values on the horizontal strip have a product equal to $1+{\cal O}(\epsilon^p)$, for some positive integer $p$. In what follows we shall refer to the exponents $p$, $q$ as `weights'. During the interactions of the singularities, quantities $\epsilon^r$ may appear (where $r$ is a nonnegative integer) and the term `weight' will be used in this case as well.}

In order to illustrate the effect of a different choice of weight we consider the simplest situation where the various infinities on the {SW/NE diagonal} are obtained as limits of a quantity proportional to $1/\epsilon$,  while the inverse values on the strip have a product equal to $1+{\cal O}(\epsilon^2)$. The result of the interaction is presented in Figure 11. 
\bigskip
\centerline{\includegraphics[width=5.cm,keepaspectratio]{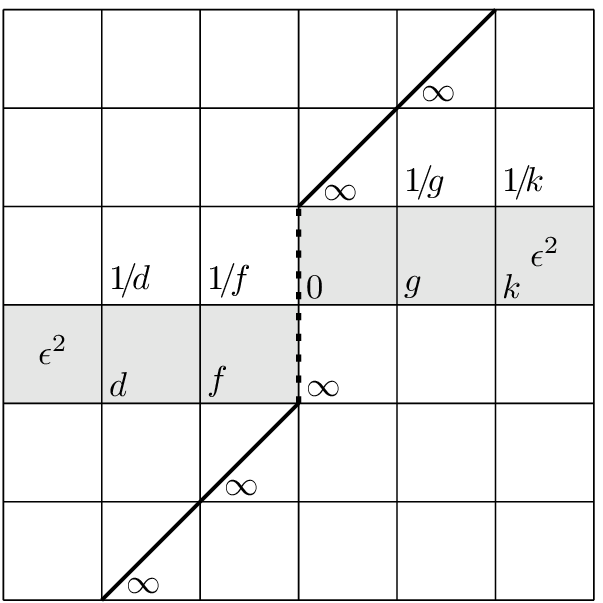}}
\vskip.25cm\centerline{~~~~{\sl Figure 11}}\vskip.5cm

The line of infinities {again jumps upwards by two lattice spacings whereas the horizontal strip is now shifted only by one spacing. }No elementary cell is formed by the interaction this time. {We would like to point out that the existence of this singularity was already surmised in our previous work [\doyong] on the basis of the results obtained for one-dimensional reductions of the discrete KdV.} 

The interaction of a strip of weight 2 with two {consecutive diagonals of infinities (each of weight 1)} leads to what would one expect from the  interaction of the strip with each line of infinities independently, as shown in Figure 12.
\bigskip
\centerline{\includegraphics[width=6cm,keepaspectratio]{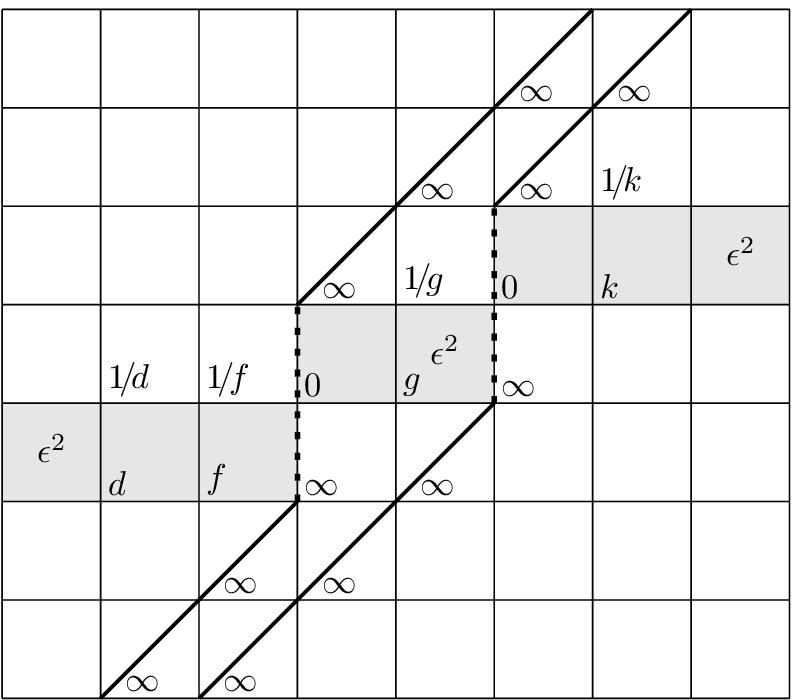}}
\vskip.25cm\centerline{~~~~{\sl Figure 12}}\vskip.5cm

{As can be assessed from Figure 13, a singularity where infinities and zeros alternate, i.e. a singularity of the third type encountered in section 2, leads to similar results.}
\bigskip
\centerline{\includegraphics[width=6cm,keepaspectratio]{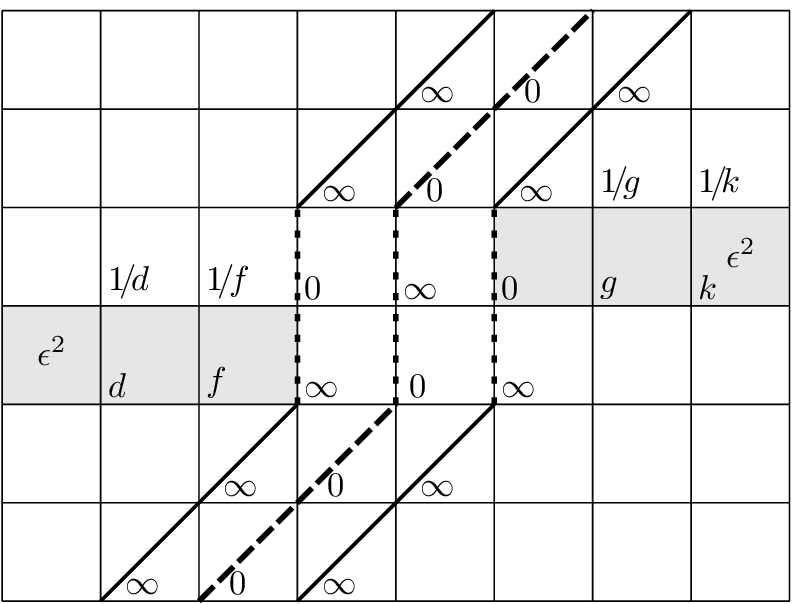}}
\vskip.25cm\centerline{~~~~{\sl Figure 13}}\vskip.5cm

{In what follows we shall not give any more results involving the latter type of singularity since they always closely resemble those obtained for the second type of singularity, without any new features appearing. Nor shall we present any results for cases where we have more than one singularity of the second type, since their interactions with a horizontal strip are all independent anyway.}

{Based on the phenomenon} depicted in Figure 11 it is natural to wonder what happens when {the ratio of the powers of $\epsilon$ that leads to the second type of singularity and those related to the horizontal strip is different from 2.} Clearly one can always redefine $\epsilon$ so as to have $q=1$ for the line of infinities and analyse the results for the possible values of $p$, however the simplest way to perform the calculations is by keeping $q$ so as to avoid fractional exponents. In Figure 14 we present the results in the case $p=3, q=1$. 
\bigskip
\centerline{\includegraphics[width=5.cm,keepaspectratio]{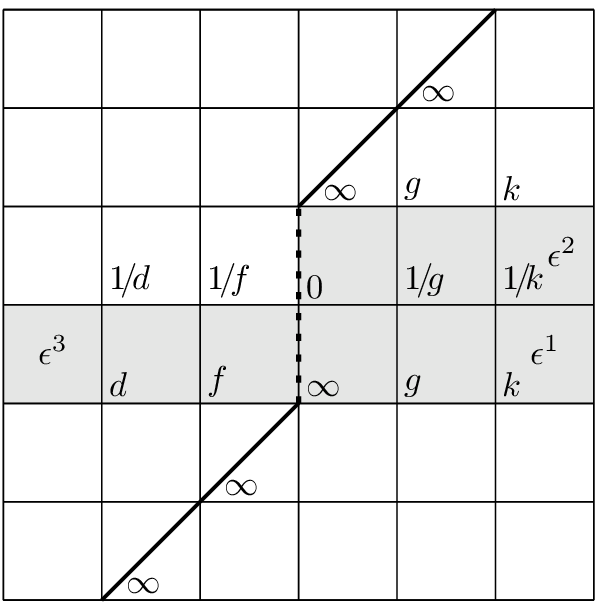}}
\vskip.25cm\centerline{~~~~{\sl Figure 14}}\vskip.5cm

{Here we discover a new phenomenon:} while the line of infinities behaves just as in the case of $p=2$, $q=1$, the {horizontal strip has now doubled in width after interaction. In the lower part of the strip, the reciprocal values have a product of weight 1, while in the upper part the weight is equal to  2.} This is the general behaviour as long as $p/q>2$: the interaction splits the strip into two parts, the upper part always having weight $2q$ while the lower part (which necessarily lies in the prolongation of the impinging strip) has weight $p-2q$. This shows that this type of singularity can exist in two forms: either one where all the weight is concentrated on a single strip or one where the weight is split over two adjacent strips. Both forms are avatars of the same singularity.

Next we turn to the case $2>p/q>1$. An example where $p=5, q=3$ is shown in Figure 15.  
\bigskip
\centerline{\includegraphics[width=5.5cm,keepaspectratio]{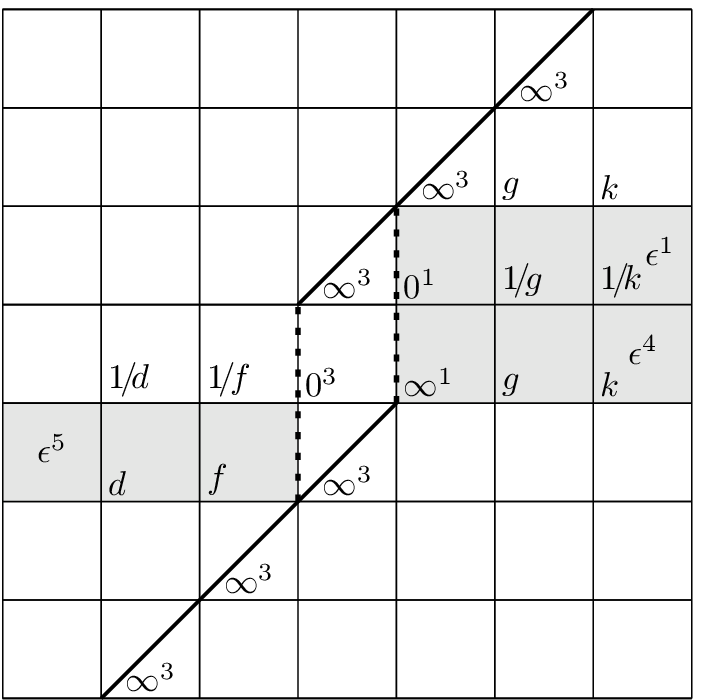}}
\vskip.25cm\centerline{~~~~{\sl Figure 15}}\vskip.5cm

{In Figure 15 we have introduced exponents corresponding to the weights of all the zeros and infinities in order to be able to follow the evolution. As before, the horizontal strip is again split into two parts, the whole of which is now pushed up by one spacing.} This is the general behaviour when $2>p/q>1$: the lower part of the strip, after interaction, is shifted upwards once with respect to the incoming strip and has weight $2(p-q)$, while the part  {stacked} on top of it has weight $2q-p$. 

The last situation we must examine is that where $p/q<1$. Here two cases must be distinguished. The first one corresponds to $p=2$ with $q>2$ ($q$ may be even). The behaviour of such a case is shown in Figure 16 for $q=4$. {The horizontal strip does {\sl not} split in this case but is pushed $q$ spacings upwards (here $q=4$).}
\bigskip
\centerline{\includegraphics[width=5.5cm,keepaspectratio]{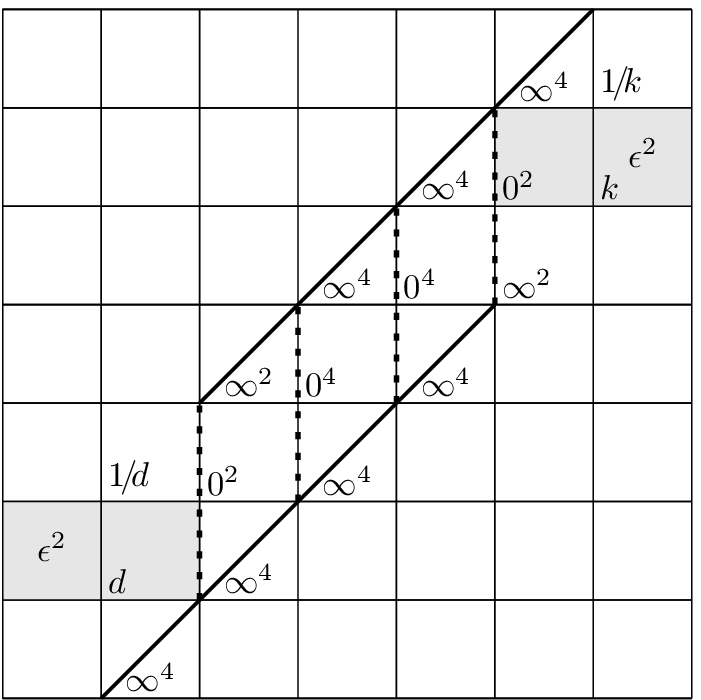}}
\vskip.25cm\centerline{~~~~{\sl Figure 16}}\vskip.5cm

When $p$ is not equal to 2 one must find {the unique integer $Q$} such that $2/Q>p/q>2/(Q+1)$. Take for instance $p=3, q=7$. In this case $Q=4$ since $2/4>3/7>2/5$. In Figure 17 we show the interaction of the corresponding singularities. 
\bigskip
\centerline{\includegraphics[width=6.cm,keepaspectratio]{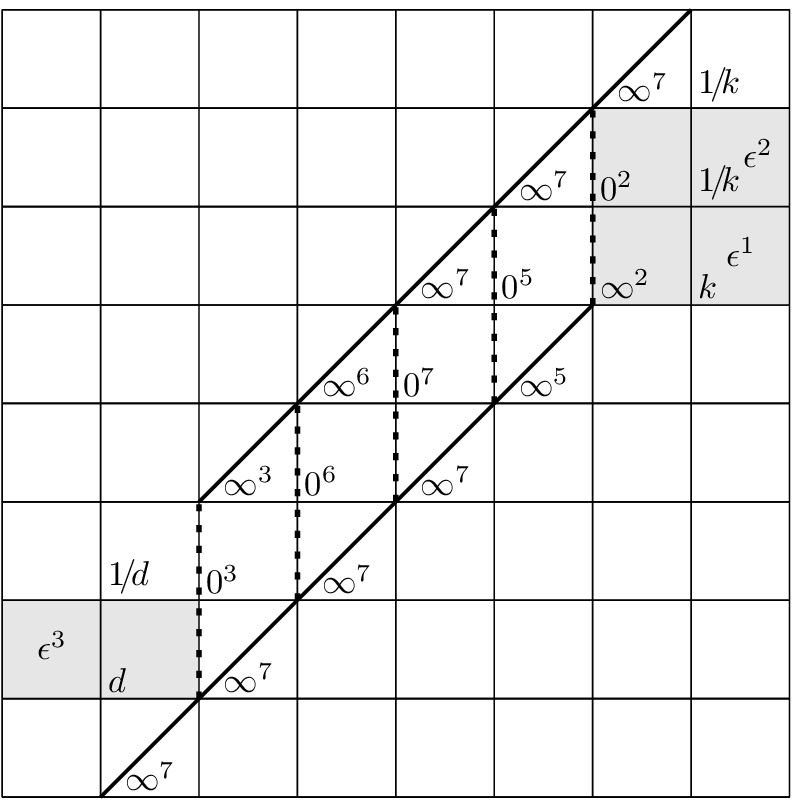}}
\vskip.25cm\centerline{~~~~{\sl Figure 17}}\vskip.5cm

In general, the strip will be split into two parts. {The lower part is pushed $Q$ spacings upwards and has weight $p(Q+1)-2q$,  while the upper part  {stacked} on top of it has weight $2q-Qp$.}

While the interactions in the cases we presented above may appear complicated, the underlying rule is very simple. The result of the interaction of {a horizontal strip with weight $p$ with a diagonal of infinities with weight $q$,  is to push the strip upwards by $k=\lsqbr{2q/p}\rsqbr$ spacings (where $\lsqbr x\rsqbr$ denotes the integer part of an real number $x$). If $2q/p$ is not an integer itself, then the strip splits into two  {adjacent} strips with weights $(k+1)p-2q$ for the lower one and $2q-kp$ for the upper one. The overall weight of the horizontal strip is obviously conserved in this process. 
It is important to stress that this rule is valid for arbitrary weights $q$ and $p$ and that it encompasses all the interactions of a strip with a diagonal of infinities we discussed until now.}

{Based on this remarkable observation we would like to introduce the notion of a {\sl\tan} to describe the singularity that underlies this phenomenon: by a \ten we mean an isolated strip or an isolated set of two contiguous strips (here horizontal but they might be vertical), the total weight of which is conserved under interaction with other singularities (where by isolated we mean separated by at least two lattice spacings from any other such strip(s)).

The term `\tan' derives from the Japanese naming of elementary particles (always ending in -shi (\begin{uCJK}子\end{uCJK})) and the character for `band' (\begin{uCJK}帯\end{uCJK}).

In several of the examples we presented above we observed a splitting of the \ten into two  {contiguous} parts. Of course, a \ten consisting of two strips is not necessarily the result of an interaction: one can imagine initial conditions corresponding to a single \ten that span two adjacent strips. It is therefore interesting to study the interaction of such a singularity with a line of infinities {to see how the above definition holds up.} Figure 18 presents such a case. 
\bigskip
\centerline{\includegraphics[width=5cm,keepaspectratio]{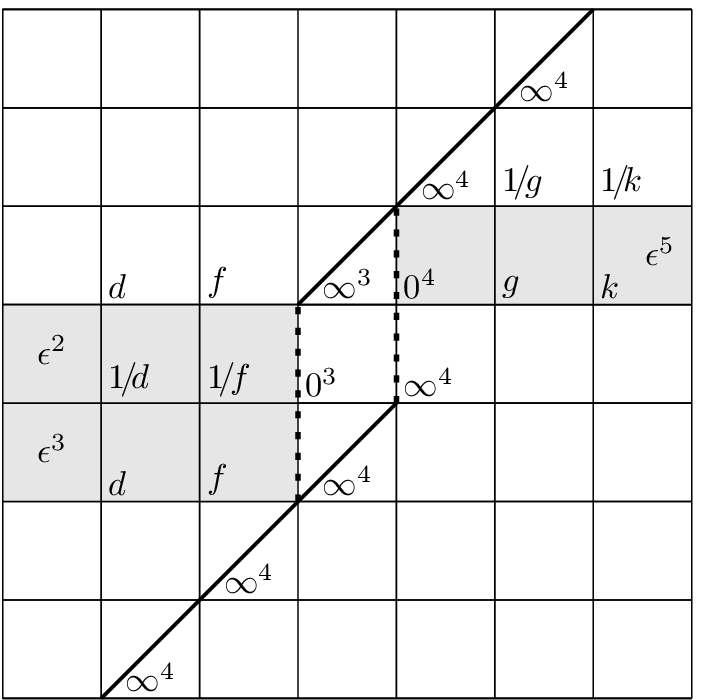}}
\vskip.25cm\centerline{~~~~{\sl Figure 18}}\vskip.5cm

Here a new phenomenon appears: the two incoming strips of the \ten merge into a single one. As we shall explain shortly, this is not an exceptional situation. A fusion of the two strips of a \ten can easily occur, depending on the weight of the \ten and that of the line of infinities. Still, it is more frequent for a \ten to re-distribute its weight over two strips as the result of an interaction, usually with an overall upshift by a small number of lattice spacings.  An example of such a situation is presented in Figure 19.  
\bigskip
\centerline{\includegraphics[width=5cm,keepaspectratio]{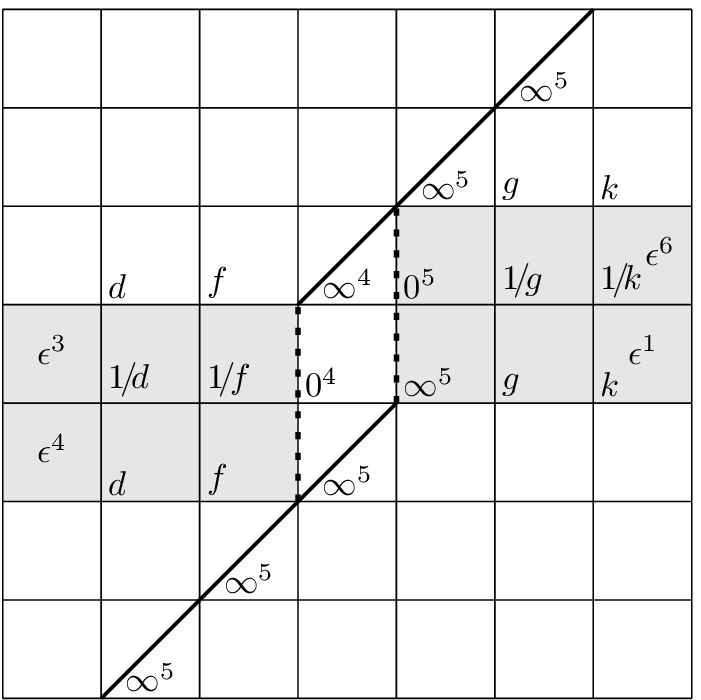}}
\vskip.25cm\centerline{~~~~{\sl Figure 19}}\vskip.5cm

The general interaction of a \ten with weights $n$ (for the lower part) and $p$ (for the upper part) with a line of infinities of weight $q$ leads to a \ten with respective weights $\tilde n, \tilde p$,  the whole shifted upwards (compared to the incoming part) by $k$ spacings, where the values of $k,\tilde n$ and $\tilde p$ are governed by the equations}
$$\tilde n+\tilde p=n+p\,,\eqdaf\zhep$$
$$k\tilde n+(k+1)\tilde p=p+2q.\eqno(\zhep b)$$
{Since all variables are required to take nonnegative integer values these equations are easily solved in terms of $n,p$ and $q$. Eliminating $\tilde n=n+p-\tilde p$ from (\zhep b) we obtain the expression
$$2q+p = k (n+p) + \tilde p,$$
which, when $\tilde n>0$ (and thus $\tilde p<n+p$) has the unique solution:
$$k=\Big[ {2q+p\over n+p}\Big]\,,\qquad \tilde p = 2q+p - (n+p) k.\eqdef\newone$$
The exceptional case $\tilde n=0$ occurs  (only) when $2q+p$ is a multiple of $n+p$, and corresponds to
$$k={2q+p\over n+p}-1\,,\qquad \tilde p = n+ p.\eqdef\newtwo$$

Note that this last case is a slightly pathological re-interpretation of the special case $\tilde p=0$ of (\zhep), for which $\tilde n=n+p$ and $k=(p+2q)/(n+p)$. Clearly, these special cases correspond to the merging of two strips into a single one, as shown in Figure 18. 

The interaction of horizontal and vertical \ten being trivial, the investigations we carried out above show that the \ten can only possibly interact among themselves through the intermediary of singularities of the second or third type.
In particular, when two \ten are close enough so that the result of their interactions with a line of infinities would have the lower one encroach upon the space occupied by the upper one, we expect something to happen. When is such an interaction possible? As explained in the previous paragraphs a \ten of weight $p$ interacting with a line of weight $q$ is pushed upwards by $\lsqbr2q/p\rsqbr$ spacings. So, if the second \tan, say with weight $P$, happens to lie less than $\lsqbr2q/p\rsqbr-\lsqbr2q/P\rsqbr+1$  spacings higher than the first one, an interaction might take place. Let us illustrate this with some examples. 

\bigskip
\centerline{\includegraphics[width=5.5cm,keepaspectratio]{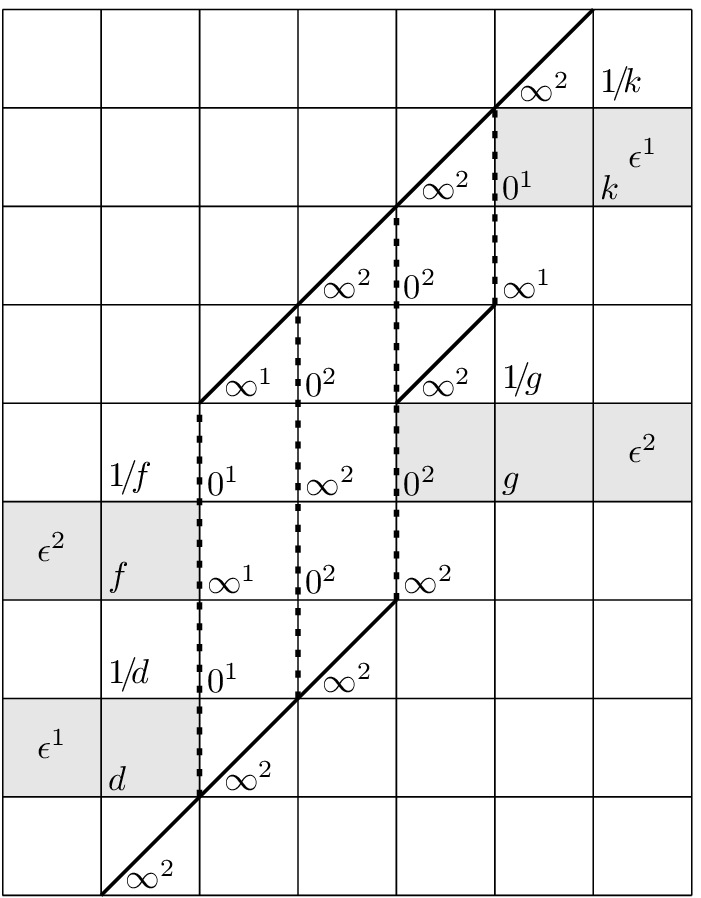}}
\vskip.25cm\centerline{~~~~{\sl Figure 20}}\vskip.5cm

In Figure 20 we show the interaction of two \ten of weight 1 and 2, respectively, with a line of weight 2. Without any interaction between the two \tan, the interaction with the diagonal would have pushed the lower \ten upwards by 4 lattice spacings and the higher one by two, such that they would have to occupy the same space. As a result there is an interaction between the two \tan: the lower one is pushed a further two lattice spacings upwards, whereas the upper one is pushed downwards by 1 lattice spacing compared to where it would have been had it interacted alone with the line. It can be checked that if one places the upper \ten at least two positions higher than in Figure 20, no interference between the two \ten occurs. 

\bigskip
\centerline{\includegraphics[width=4.cm,keepaspectratio]{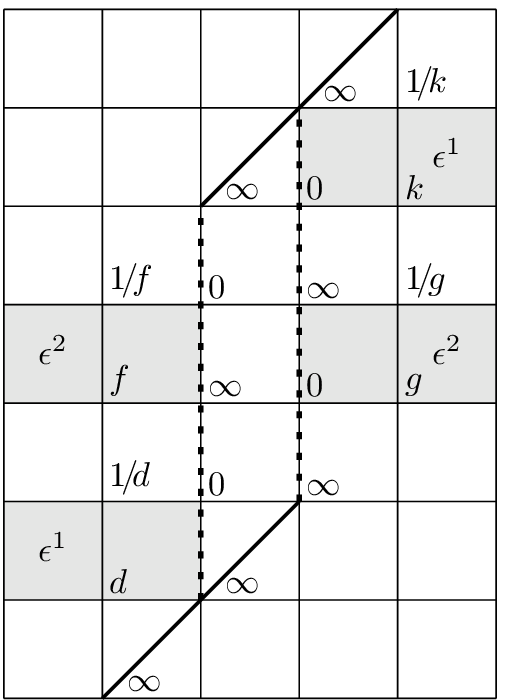}}
\vskip.25cm\centerline{~~~~{\sl Figure 21}}\vskip.5cm
On the other hand, as shown in Figure 21, taking the same two \ten but having them interact with a line of weight 1 results in the lower \ten ending up two spacings lower (and the upper one, one spacing lower) than in Figure 20, which is exactly the difference in the upshifts after interaction with the line of infinities,  induced by the change in its weight, $q$. This suggests that the extra shifts stemming from the interference among the \ten are in fact independent of $q$. Note also that without the interaction between the two \tan, their interaction with the diagonal of infinities would have put them not in the same position on the lattice (as was the case for the situation in Figure 20) but at vertically adjacent positions. This shows that the conditions under which an interaction between \ten takes place are rather subtle.

In Figure 22 we present a slightly more complicated situation in which the interaction splits both \tan. 

\bigskip
\centerline{\includegraphics[width=5cm,keepaspectratio]{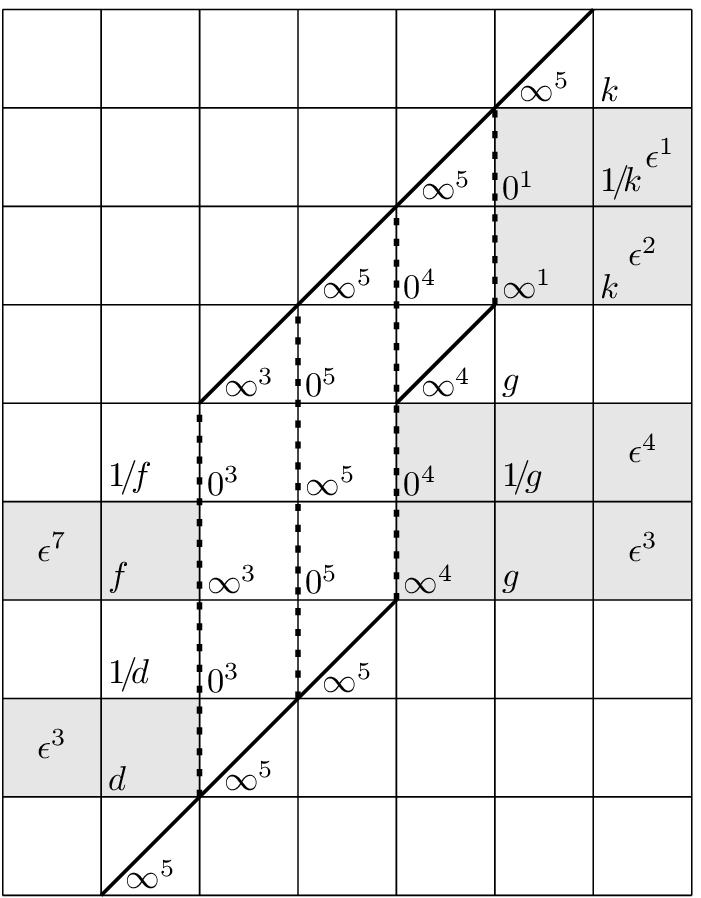}}
\vskip.25cm\centerline{~~~~{\sl Figure 22}}\vskip.5cm

In the absence of interference between the two \ten the result would have been a splitting of the lower \ten into strips of weight 2 and 1 respectively (counting from the bottom) the whole pushed upwards three spacings, while the upper \ten would have been pushed one spacing upwards after having split into two  {contiguous} strips of weights 4 and 3 respectively (again counting from the bottom). This would have put both \ten at exactly the same position, hence the interaction between the two. This interaction now shifts the lower \ten upwards by a further two lattice spacings, as before, but what happens to the upper one is more interesting: it is not upshifted at all and splits into two  {contiguous} strips of weights 3 and 4 respectively (counting from the bottom), exactly opposite to what one would expect after interaction with a diagonal of weight 5.

Again this complicated looking behaviour can be explained by a simple interaction rule. 
As we saw above,  interaction with a diagonal of infinities with weight $q$ of a  \ten with weights $n$ and $p$ (counting from the bottom, and where $p$ might be 0)  yields a \ten with weights $ \tilde n= n+ p-\tilde p,\, \tilde p= 2q+p - k(n+p)$ where $k=\Big[{2q+p\over n+p}\Big]$ is the upshift incurred by the entire \tan. If we now have a second \tan, with respective weights $N$ and $P$, placed some distance above the first one, there are three possible interaction scenarios.

[A] If the weights involved and the distance between the \ten are such that the result of the interaction of the lower one with the diagonal is still positioned below the result of the interaction of the upper one with the diagonal, such that these are well-separated (i.e. at least by one lattice spacing), then the \ten do not interact among themselves and the net result is just an $\tilde n= n+ p-\tilde p,\, \tilde p= 2q+p - k(n+p)$ \ten upshifted $k=\Big[{2q+p\over n+p}\Big]$ lattice spacings with respect to its original position, accompanied by an $\tilde N= N+ P-\tilde P,\, \tilde P= 2q+P - K(N+P)$ \ten upshifted $K=\Big[{2q+P\over N+P}\Big]$ lattice spacings. Note that this tells us that if the lower \ten is heavier than the upper one the \ten never interact.

[B] If the following sets of horizontal strips
\par$\bullet$ two strips with respective weights $ \tilde n= n+ p-\tilde p$ and $\tilde p= 2q+p - k(n+p)$, $k=\Big[{2q+p\over n+p}\Big]$, upshifted $k+2$ lattice spacings with respect to the position of the $n,p$ \ten
\par$\bullet$ two strips $\tilde N= N+ P-\tilde P,\, \tilde P= 2(q-n-p)+P - K'(N+P)$ upshifted $K'=\Big[{2q+P-2(n+p)\over N+P}\Big]$ lattice spacings with respect to the position of the $N,P$ \ten
\par  are separated by at least one lattice spacing, {then the interaction between the two \ten has been completed and} the above expressions give the new forms (and positions) of the \ten that appear as the net result of their interaction with the diagonal of infinities and the interaction among themselves. Figure 23 is an example of such an interaction with $q=5$, $n=1, p=2$ and $N=5, P=2$.

\bigskip
\centerline{\includegraphics[width=5cm,keepaspectratio]{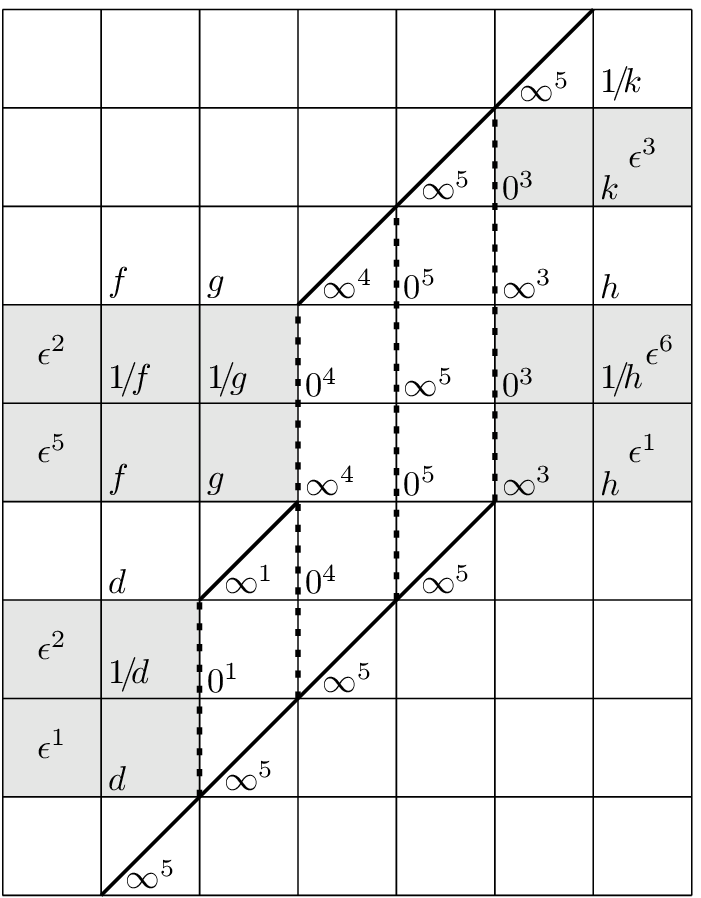}}
\vskip.25cm\centerline{~~~~{\sl Figure 23}}\vskip.5cm

Note that scenario [B] is incompatible with scenario [A] but that it is possible that an interaction yields \ten that are not well separated, whereby violating both the conditions for scenario [A] as well as those for [B].
{The result of this third possible scenario however, in which the interaction is still ongoing, is highly complicated and might, in all generality, even be impossible to formulate in a `closed form' such as those given for scenarios [A] or [B].}
However, there exists an iterative  procedure by which this type of interaction (and actually all other types as well) can be obtained.

\bigskip
4. {\scap Symbolic dynamics for taishi interactions}
\medskip
In all the examples presented in the previous paragraphs the evolution of the taishi was rather simple: one could follow the taishi from pre- to post-interaction. However this is (by far) not always the case. For example, in Figure 24 two taishi interact with a line of weight 3, resulting in a structure which occupies four adjacent strips and in which one can no longer distinguish the individual taishi. 
\bigskip
\centerline{\includegraphics[width=5cm,keepaspectratio]{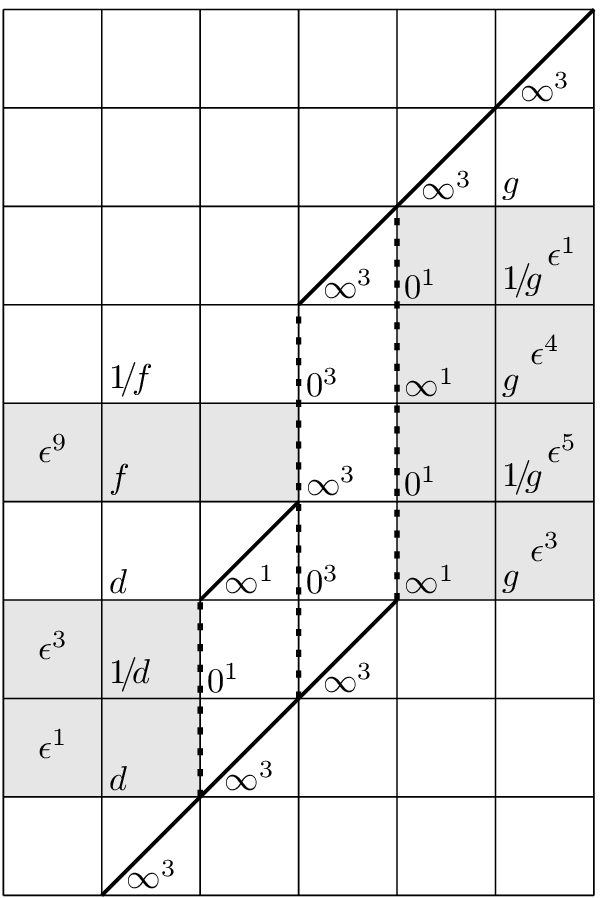}}
\vskip.25cm\centerline{~~~~{\sl Figure 24}}\vskip.5cm
However, as we show in Figure 25, one more interaction of this quadruple strip with a diagonal line with a well-chosen weight suffices for the taishi to emerge from the interaction well separated. 
\bigskip
\centerline{\includegraphics[width=5cm,keepaspectratio]{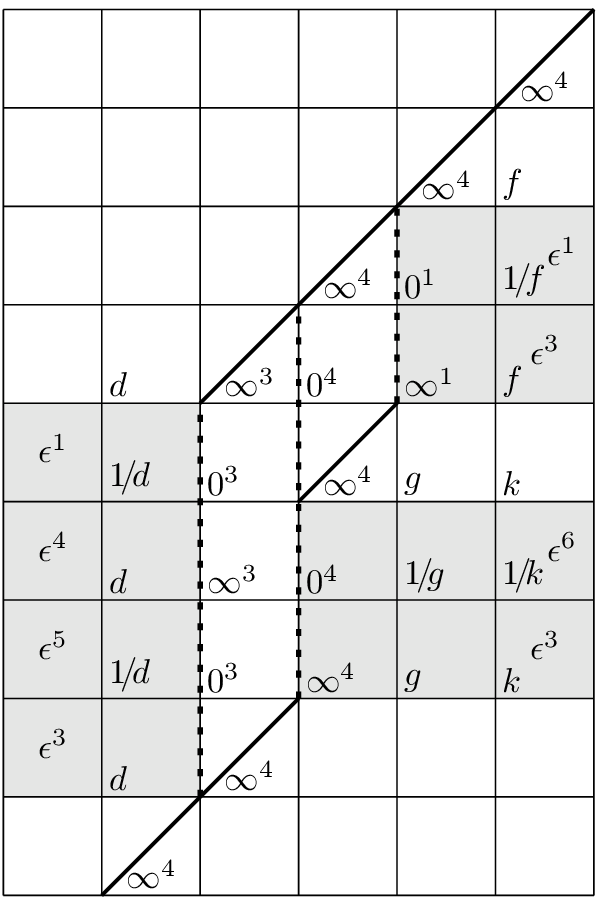}}
\vskip.25cm\centerline{~~~~{\sl Figure 25}}\vskip.5cm

By choosing a diagonal line of weight 4 one sees that the taishi of weight 9 re-emerges, albeit split into two parts with weights 3 and 6, while the taishi of {weight 4 (split into 3 and 1)} emerges well above it. 
Providing rules for such interactions in the spirit of what was done in the previous section looks, a priori, prohibitively difficult. However, as we shall explain in a moment, nothing is further from the truth. In fact, the interactions presented till now and the rules governing them can be summarised in a very simple way.

In section 3 we already pointed out (in connection to Figures 9 and 12) that successive interactions of a \ten with a diagonal of infinities are actually independent. In fact, it is not hard to see that the net effect on a taishi of such successive interactions with  a diagonal of infinities is just a cumulative one: if a \ten interacts with diagonals of weigths $q_i\, (i=1,\dots, \ell)$ (in any order), after the last interaction the net effect on the taishi is equal to that of a single interaction with a diagonal of weight $q_1+\cdots+q_\ell$. Take for example the case of a single \ten with weights $n$ (for the band on the bottom) and $p$ (for that on top), that interacts with two diagonals of weights $q$ and $\tilde q$, in that order. Formula (\newone) tells us that after the first interaction we will have a \ten with weights $\tilde n= n+ p-\tilde p$ and $\tilde p= 2q+p - k(n+p)$, upshifted $k$ lattice spacings, where $k=\Big[{2q+p\over n+p}\Big]$. When, in turn, this \ten interacts with the diagonal of weight $\tilde q$, the accumulated result of both interactions yields a \ten with weights $\hat n$ and $\hat p$, upshifted $k + \Big[{2\tilde q + \tilde p\over \tilde n+ \tilde p}\Big] = \Big[{2(\tilde q + q) + p\over n+ p}\Big]$ lattice spacings with respect to the position of the original weight $n,p$ \tan, where 
$$\hat p = 2 (q+\tilde q) + p - \Big[{2(\tilde q + q) + p\over n+ p}\Big] (n+p)\,,\quad \hat n = n+ p - \hat p,$$
i.e. exactly the result of an interaction of the original \ten with a diagonal of weight $q+\tilde q$.

This allows us to describe the general interaction of a \ten with a diagonal of weight $q$, for example,  as $2q$ successive interactions of that \ten with a diagonal of weight $\miso$ instead. The reason we pick a diagonal of $\miso$ is that such an interaction turns out to be elementary.

As already noted in section 3, by rescaling the respective weights, one can re-interpret the interaction of an $n,p$ \ten with a diagonal of weight $\miso$ as that of a $2n,2p$ \ten with a diagonal of weight 1: the induced shifts on the lattice will be equal, $k=\Big[{p+1\over p+n}\Big]$, and the weights of the resulting \ten are the double of those of an $n,p$ one, i.e. ultimately one has:
$$\tilde n = n -1 + k (n+p)\,,\qquad\tilde p = p +1 - k(n+p).$$
Elementary arithmetic then shows that when $n>1$ one has $k=0$ and $\tilde n=n-1,\, \tilde p=p+1$, whereas when $n=1$ one has $\tilde n = p+1,\, \tilde p = 0$ and when $n=0$, $\tilde n=p-1$ and $\tilde p=1$, both upshifted 1 position compared to the original \tan. This result can be summarized very easily: if $n=0$ the resulting \ten is obtained by subtracting 1 from the strip with weight $p$ and adding 1 to the weight of the strip just above it; whereas when $n\geq1$ the resulting \ten is obtained by subtracting 1 from the weight in the bottom strip (with weight $n$) and adding 1 to that of the strip above it.

It turns out that this extremely simplified dynamics is valid for any configuration of horizontal strips. This leads to the following simple algorithm for calculating the interaction of a collection of horizontal strips with various weights (some of which can be 0) with a diagonal line of weight $q$.

{\sl Starting sufficiently low,

- move upwards up to the first (i.e. lowest) horizontal strip with non-zero weight and subtract 1 from its weight and add 1 to the weight of the strip just above it, 

- then move to the next horizontal strip with non-zero weight above those two strips, subtract 1 from its weight and add 1 to the weight of the strip just above it,

- repeat the same procedure until there are no more strips with non-zero weight left.

The resulting collection of strips is that obtained from interaction with a diagonal of weight $\miso$ ; to obtain the result of an interaction with a diagonal of weight $q$ repeat this procedure an extra $2q-1$ times.}

We can now illustrate this rule by applying it to a realistic situation. As such we choose the taishi interactions presented in Figures 24 and 25. We give below the evolution of the weights, using boldface numbers for the results of iterations with an even number of steps i.e. the ones corresponding to integer values of $q$. Since the weight of the diagonal in Figure 24 is 3 we will need 6 steps in order to obtain the result shown in the figure. We find successively
$$\matrix{\bf 0\cr\bf0\cr\bf9\cr\bf0\cr\bf3\cr\bf1}\Longrightarrow\matrix{0\cr1\cr8\cr0\cr4\cr0}\Longrightarrow
\matrix{\bf 0\cr\bf2\cr\bf7\cr\bf1\cr\bf3\cr\bf0}\Longrightarrow\matrix{0\cr3\cr6\cr2\cr2\cr0}\Longrightarrow
\matrix{\bf 0\cr\bf4\cr\bf5\cr\bf3\cr\bf1\cr\bf0}\Longrightarrow\matrix{0\cr5\cr4\cr4\cr0\cr0}\Longrightarrow\matrix{\bf 1\cr\bf4\cr\bf5\cr\bf3\cr\bf0\cr\bf0}$$
and remark that after six steps we obtain precisely the weights of the strips on the right-hand side of Figure 24. Next we start from the strips of Figure 25
$$\matrix{\bf0\cr\bf0\cr\bf 1\cr\bf4\cr\bf5\cr\bf3}\Longrightarrow\matrix{0\cr0\cr2\cr3\cr6\cr2}\Longrightarrow
\matrix{\bf0\cr\bf0\cr\bf 3\cr\bf2\cr\bf7\cr\bf1}\Longrightarrow\matrix{0\cr0\cr4\cr1\cr8\cr0}\Longrightarrow
\matrix{\bf0\cr\bf1\cr\bf 3\cr\bf2\cr\bf7\cr\bf0}\Longrightarrow\matrix{0\cr2\cr2\cr3\cr6\cr0}\Longrightarrow
\matrix{\bf0\cr\bf3\cr\bf 1\cr\bf4\cr\bf5\cr\bf0}\Longrightarrow\matrix{0\cr4\cr0\cr5\cr4\cr0}\Longrightarrow\matrix{\bf1\cr\bf3\cr\bf 0\cr\bf6\cr\bf3\cr\bf0}$$
and again obtain the weights of the emerging taishi after 8 steps.

A final example of a complex series of interactions is given in Figure 26.
\bigskip
\centerline{\includegraphics[width=7cm,keepaspectratio]{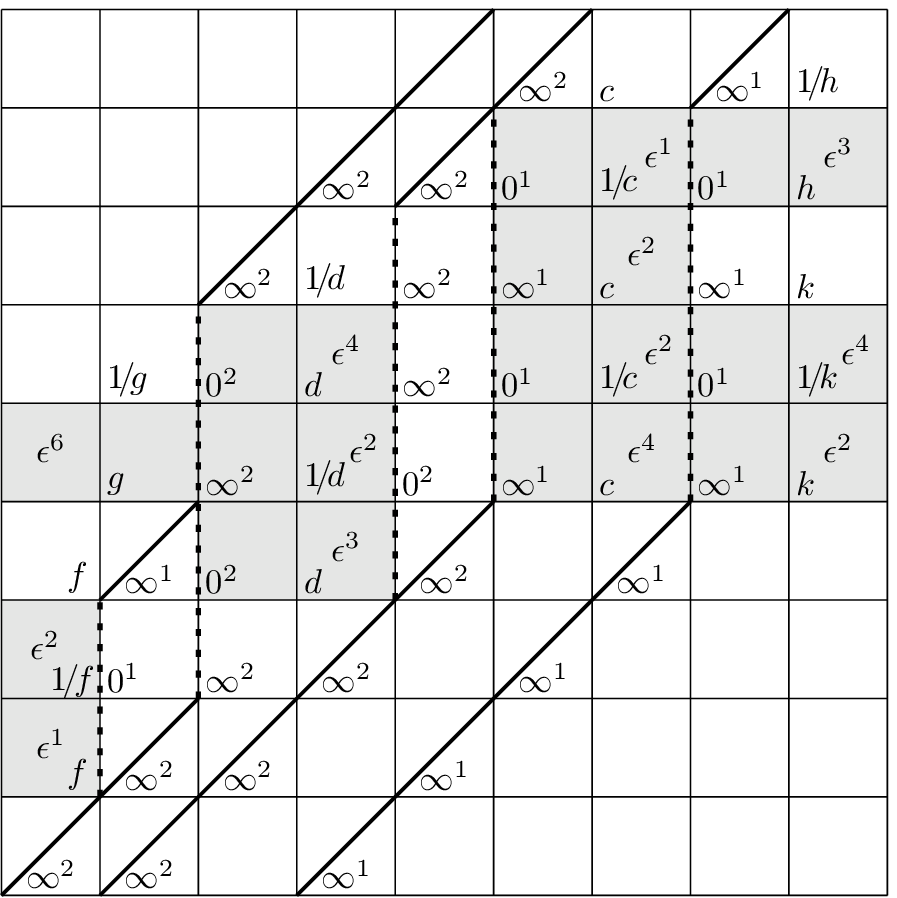}}
\vskip.25cm\centerline{~~~~{\sl Figure 26}}\vskip.5cm
The impinging taishi have weights 3 (split as 1+2) and 6 respectively. After three interactions with diagonal lines of weights 2, 2, and 1 they emerge with the taishi of weight 6 split as 2+4 and that of weight 3 lying above the former, and well separated from it.

The interpretation in terms of the taishi sybolic dynamics we introduced above is straightforward. 
$$\matrix{\bf0\cr\bf0\cr\bf0\cr\bf 6\cr\bf0\cr\bf2\cr\bf1}\Longrightarrow\matrix{0\cr0\cr1\cr5\cr0\cr3\cr0}\Longrightarrow
\matrix{\bf0\cr\bf0\cr\bf2\cr\bf 4\cr\bf1\cr\bf2\cr\bf0}\Longrightarrow\matrix{0\cr0\cr3\cr3\cr2\cr1\cr0}\Longrightarrow
\matrix{\bf0\cr\bf0\cr\bf4\cr\bf 2\cr\bf3\cr\bf0\cr\bf0}\Longrightarrow\matrix{0\cr1\cr3\cr3\cr2\cr0\cr0}\Longrightarrow
\matrix{\bf0\cr\bf2\cr\bf2\cr\bf 4\cr\bf1\cr\bf0\cr\bf0}\Longrightarrow\matrix{0\cr3\cr1\cr5\cr0\cr0\cr0}\Longrightarrow
\matrix{\bf1\cr\bf2\cr\bf2\cr\bf 4\cr\bf0\cr\bf0\cr\bf0}\Longrightarrow\matrix{2\cr1\cr3\cr3\cr0\cr0\cr0}\Longrightarrow
\matrix{\bf3\cr\bf0\cr\bf4\cr\bf 2\cr\bf0\cr\bf0\cr\bf0}$$
It is easy to verify that after 4, 8, and 10 iterations we obtain the weights corresponding to the strips present in Figure 26. 

Clearly, one can imagine ever more complicated situations, with strips of various widths and weights interacting with a line of infinities and as a result interfering with each other, post-interaction. However the examples already shown will probably suffice to convince the reader of the impressive richness that can be found in the structure of the singularities of the lattice KdV equation.
\bigskip
5. {\scap The nonintegrable case}
\medskip
An attempt at generalising the form (\zpen) of the discrete KdV,  even ever so slightly, results into a non-integrable system. As shown in [\sincon] and again in [\doyong] the simplest generalisation of d-KdV
$$x_{m+1,n+1}=x_{m,n}+{1\over x_{m+1,n}}-{\lambda\over x_{m,n+1}},\eqdef\zoct$$

with $\lambda\ne1$, is not integrable and does not possess the singularity confinement property. 

Studying the singularities of (\zoct) is straightforward. First we remark that due to the different coefficients of the two right-most terms there exist no singularities of the \ten  type, which lie at the origin of the very rich structure of the singularities of d-KdV. On the other hand, the second and third types of singularities encountered in section 2, which consist of infinitely long parallel lines of infinities alternating (or not) with zeros and which run from south-west to north-east, do exist. {However, as noted before, although these singularities are of infinite extent in both directions, they are not considered as being unconfined.}

The violation of confinement {in this nonintegrable case} comes from the singularities of the first type. If a zero appears at some point on the lattice due to a particular choice of initial condition, the singularities this creates do not disappear after a few iteration steps but {persist} indefinitely. Figure 27 gives two such examples, one for a single zero, in the upper left part of the figure, and one for two adjacent zeros, in the lower right part.
\bigskip
\centerline{\includegraphics[width=7cm,keepaspectratio]{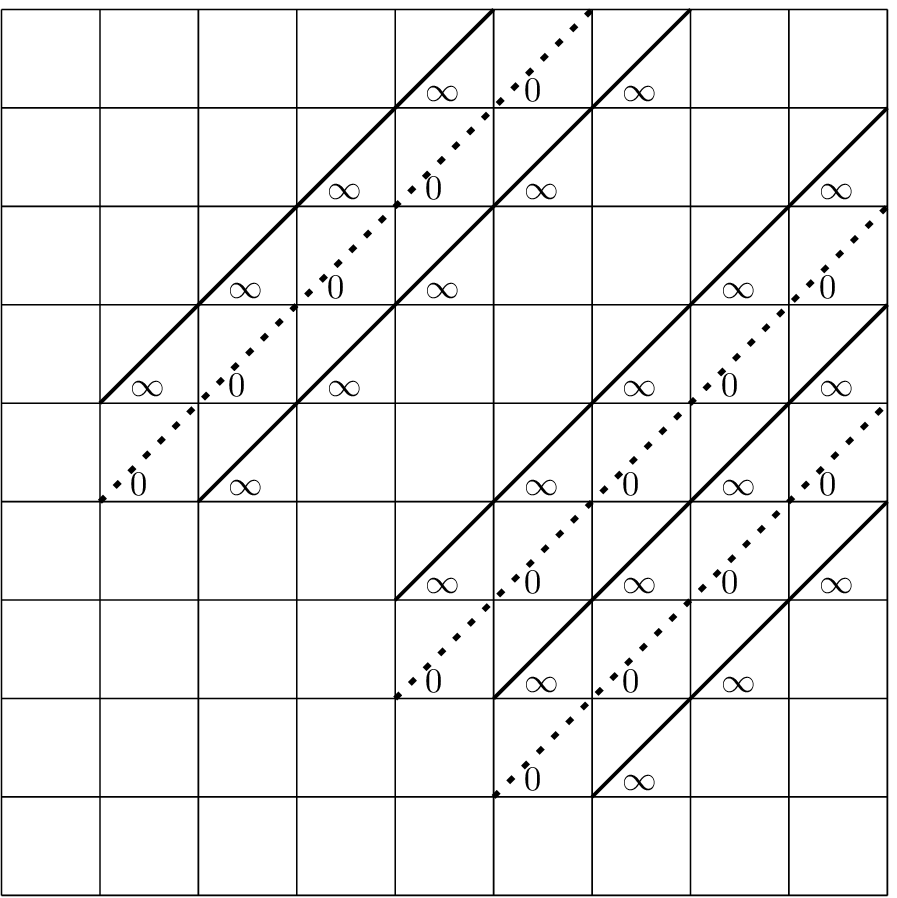}}
\vskip.25cm\centerline{~~~~{\sl Figure 27}}\vskip.5cm

Clearly these unconfined singularities can be combined with the infinite lines of infinities and/or infinities alternating with zeros. For instance, if a zero appears spontaneously next to an infinite line of infinities, the latter will not be perturbed while the consequence of the zero will be to create a line of zeros (and a line of infinities on its left) extending unimpeded towards the north-east. 
The signature of an unconfined singularity is the existence of (at least) one line of zeros, starting at some point on the lattice and proceeding unhindered in the north-east direction but not extending both ways, accompanied by lines of infinities on its left and its right. One of these lines {of infinities} (but not both) might extend to infinity in both directions. 
\bigskip
6. {\scap Summary and conclusions}
\medskip
In this paper we set out to study the singularities of the discrete KdV equation, {which is a question we already addressed in [\doyong].} However, in that paper we limited ourselves to the study of a single type of singularity, namely those singularities that arise due to a zero appearing at a certain point on the lattice because of a particular choice of initial conditions (although the fact that contiguous zeros can lead to more complicated singularities was also recognised). Most importantly, singularities due to one or more zeros disappear after some iteration step, a fact that led to the discovery of the singularity confinement property [\sincon]. The singularities of d-KdV were not studied any further in [\sincon] since the integrability constraints can be obtained by the study of the simplest singularity pattern. One point however that was not explicitly stressed in [\sincon], is that  a singularity consisting of a finite number of adjacent zeros is confined in a finite number of steps.  

Given the recent progress in the characterisation of singularities of discrete systems, with the discovery of the role played by cyclic or anticonfined singularities, it became interesting to examine afresh the singularity structure of d-KdV. A first attempt was that of [\doyong] where we studied the singularities of a family of reductions of d-KdV and inferred from it the possible structure of the singularities of the full, non-reduced, system. However, studying one particular type of reduction cannot reveal all the richness of the initial system: this would necessitate the study of all possible reductions or at least a sufficiently large number of them allowing to make educated guesses concerning the full system. It was thus clear that the problem had to be tackled head-on.

What was done in this paper was first to identify the possible singularities of the discrete KdV. This resulted in four different types. The first type are the localised, confined, singularities already identified in [\sincon]. The three remaining types of singularities are of infinite extent ({a property which is not necessarily} incompatible with integrability). The second and third type consist of diagonals of infinities going in the south-west to north-east direction, for the second type, while for the third type these lines of infinities alternate with lines of zeros. The fourth type of singularity are horizontal (vertical) strips, single or multiple, in which the product of two vertically (horizontally) adjacent values is 1 ($-1$). This fourth type of singularity is particularly interesting since it can interact with singularities of the other three types, {allowing us to systematically study and, ultimately, make sense of the incredibly rich nature of the singularity patterns for d-KdV.} We were therefore compelled to invent a new name for this type of singularity, opting for the term `\tan', inspired by the Japanese word for `band'. The \ten obey a very elementary type of interaction rule, when interacting with singularties of the second type. This led us to formulate a symbolic algorithm that yields the outcome of the interaction of a general configuration of \ten with a singularity of type two. Besides the elementary nature of the interactions between singularities of the fourth type (=\tan) another noteworthy feature of their interaction properties is the `fixed shift' the lightest  \ten acquires after a mixed interaction with a type-two singularity and another \tan. This shift by two lattice spacings is
eerily reminiscent of the phase-shift incurred by the background in the interaction with solitons in the ultradiscrete KdV equation [\refdef\udKdV].

It is the absence of \tan-type singularities in the non-integrable case that leads to the relative paucity of singularity structures in this case. The only thing that happens in the non-integrable case is that the confined singularities of the first type for the integrable d-KdV become unconfined.

Having discovered the (admittedly unexpected) richness of the singularity structure of the discrete Korteweg-deVries equation, one may wonder whether similar situations exist for other integrable lattice equations. This is a question we intend to address in future works of ours. 
\bigskip\vfill\eject
{\scap Acknowledgements}
\medskip
RW would like to acknowledge support from the Japan
Society for the Promotion of Science (JSPS), through JSPS grant
number 18K03355.

\bigskip
{\scap References}
\medskip
\begin{description}
\item{[\painleve]} P. Painlev\'e, Acta Math. 25 (1902) 1.
\item{[\ars]} M.J. Ablowitz, A. Ramani and H. Segur, Lett. Nuov. Cim. 23 (1978) 333.
\item{[\sincon]} B. Grammaticos, A. Ramani and V. Papageorgiou, Phys. Rev. Lett. 67 (1991) 1825.
\item{[\desoto]} T. Mase, R. Willox, B. Grammaticos and A. Ramani, Proc. Roy. Soc. A 471 (2015) 20140956.
\item{[\redemp]} A. Ramani, B. Grammaticos, R. Willox, T. Mase and M. Kanki, J. Phys.: Math. Theor.
 A 48 (2015) 11FT02.
\item{[\express]} A. Ramani, B. Grammaticos, R. Willox and T. Mase, J. Phys. A: Math. Theor. 50 (2017) 185203.
\item{[\halburd]} R.G. Halburd , Proc. R. Soc. A 473 (2017) 20160831.
\item{[\anticon]} T. Mase, R. Willox, B. Grammaticos and A. Ramani, J. Phys. A: Math. Theor.  51 (2018) 265201.
\item{[\abs]} V.E. Adler, A.I. Bobenko and Yu.B. Suris, Comm. Math. Phys. 233 (2003) 512.
\item{[\atkin]} J. Atkinson, SIGMA 7 (2011) 073.
\item{[\kinson]} J. Atkinson and N. Joshi, IMRN 2013 (2012) 1451.
\item{[\hirota]} R. Hirota, J. Phys. Soc. Jpn. 43 (1977) 1424.
\item{[\doyong]} D. Um, R. Willox, B. Grammaticos and A. Ramani, J. Phys. A 53 (2020) 114001.
\item{[\udKdV]} R. Willox, Y. Nakata, J. Satsuma, A. Ramani, B. Grammaticos, J. Phys. A: Theor. Math. 43 (2010) 482003.

\end{description}

\end{document}